\documentclass[a4paper]{article}
\usepackage{INTERSPEECH2019}
\usepackage{graphicx,url}
\usepackage{amssymb}
\usepackage{lineno}
\usepackage{amsmath,phonetic}
\usepackage{textcomp,algorithmic,epsfig,multirow,hhline}
\usepackage{amsfonts}
\usepackage[table]{xcolor}
\usepackage{floatrow}

\title{Speech Enhancement with Wide Residual Networks in Reverberant Environments}
\name{Jorge Llombart, Dayana Ribas, Antonio Miguel, Luis Vicente, Alfonso Ortega, Eduardo Lleida}
\address{ViVoLab, Aragon Institute for Engineering Research (I3A)\\
\textit{University of Zaragoza, Spain}
}
\email{\{jllombg,dribas,amiguel,lvicente,ortega,lleida\}@unizar.es}

\begin{document}

\maketitle
\begin{abstract}
This paper proposes a speech enhancement method which exploits the high potential of residual connections in a Wide Residual Network architecture. 
This is supported on single dimensional convolutions computed alongside the time domain, which is a powerful approach to process contextually correlated representations through the temporal domain, such as speech feature sequences. 
We find the residual mechanism extremely useful for the enhancement task since the signal always has a linear shortcut and the non-linear path enhances it in several steps by adding or subtracting corrections.
The enhancement capability of the proposal is assessed by objective quality metrics evaluated with simulated and real samples of reverberated speech signals. 
Results show that the proposal outperforms the state-of-the-art method called WPE, which is known to effectively reduce reverberation and greatly enhance the signal.
The proposed model, trained with artificial synthesized reverberation data, was able to generalize to real room impulse responses for a variety of conditions (e.g. different room sizes, $RT_{60}$, near \& far field).
Furthermore, it achieves accuracy for real speech with reverberation from two different datasets. 
\end{abstract}

\section{Introduction}
%
The high capability of the deep learning approaches for discovering underlying relations on the data has been exploited for speech enhancement tasks. 
Many interesting solutions for modeling the relationship between corrupted and clean data have been recently proposed based on a variety of DNN architectures. 
Convolutional Neural Network (CNN) based architectures have shown to effectively deal with the corrupted speech signal structure \cite{Fu2016, Park2017}.
Also, solutions based on Recurrent Neural Networks (RNN) architectures and the associated Long Short-Term Memory (LSTM) alternative have effectively been able to handle noisy and reverberant corrupted speech \cite{Maas2012,Weninger2015,Chen2016,Kinoshita2017,Gao2018}.
Both, convolutional and recurrent networks, have also appeared combined with residual blocks to further model the dynamic correlations among consecutive frames \cite{Chen2017,HanZhao2018,Santos2018}.
Residual connections make use of shortcut connections between neural network layers, allowing to handle deeper and more complicated neural network architectures, with fast convergence and a small gradient vanishing effect \cite{Zagoruyko2017}.
In this way, they are able to provide more detailed representations of the underlying structure of the corrupted signal. 

This paper proposes a novel speech enhancement method based on the Wide Residual Neural Networks (WRN) architecture using single dimensional convolutional layers.
This approach deals with reverberation in the spectral domain, making a regression from the log magnitude spectrum of reverberant speech to that of clean speech. 
In this way, it reinforces the importance of the low energy bands of the spectrum in the analysis, which have an impact on the perception of speech.  
%
%
We analyze the method performance through speech quality metrics from two viewpoints, namely the dereverberation level, and the spectral distortion introduced by the enhancement process. 
We compare the proposal performance with the state-of-the-art method called WPE in an experimental framework inspired by the REVERB Challenge task.
This method is based on the LSTM architecture and has reported top performances in this framework \cite{Kinoshita2017}.   
So far, residual connections have been barely exploited for speech enhancement. 
In \cite{Chen2017}, the authors proposed an architecture using LSTM, and they briefly studied the performance of residual connections through testing different configurations in the framework of speech recognition.
In \cite{Santos2018}, authors also proposed an architecture based on the recurrent approach but using gated recurrent units.
This study also reports results in quality measures to assess the dereverberation. 
%
%
However, they use metrics associated to PESQ, which is actually not recommended as a metric for enhanced or reverberant speech \cite{PESQ}.
%
%
In this line, our paper contributes to study the role of residual connections in deep speech enhancement solutions, assessing alternative conditions to previous studies. 
Section \ref{sec:speechenhancement} presents the proposal based on the WRN architecture, introducing the characteristics that make it interesting for speech enhancement.
Section \ref{sec:exp} describes the experimental setup. Section \ref{sec:res} shows results and discussion. 
Finally section \ref{sec:conc} concludes the paper.

\section{Proposal}
\label{sec:speechenhancement} 

\begin{figure*}[h!]
\centerline{\includegraphics[width=0.80\linewidth]{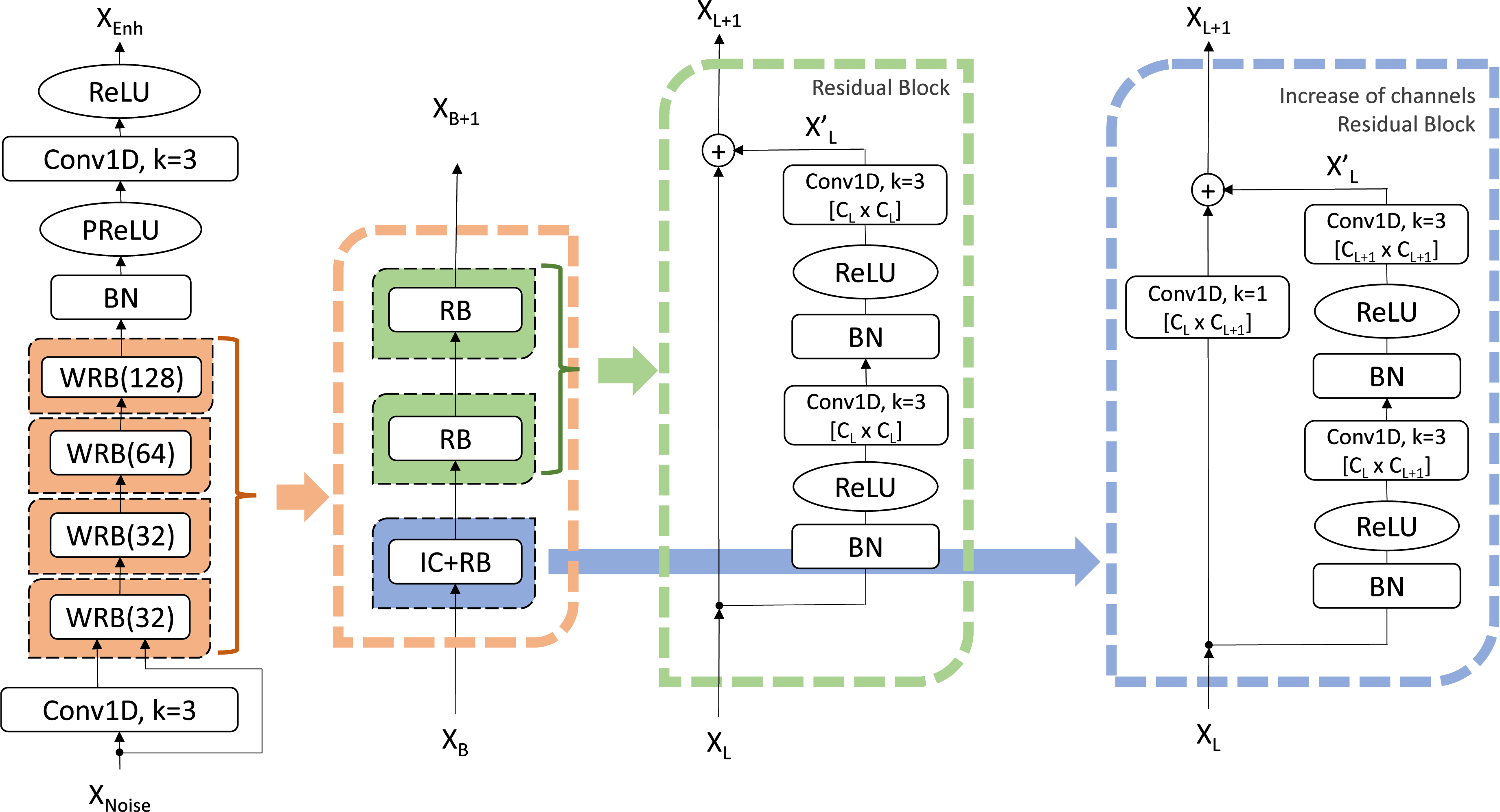}}
\caption{{\it WRN architecture proposed. From left to right there is the composition of the network blocks, with $C_{L}$ the number of channels in the layer $L$.}}
\label{fig:WRNDAE4}
\end{figure*}


%
The network architecture proposed (Figure \ref{fig:WRNDAE4}) processes input features with a first convolutional layer followed by four Wide Residual Blocks (WRB). 
The first WRB processes the output of the first convolutional layer and also its input.
Following the WRBs there is a Batch Normalization (BN) stage and a non-linearity (PReLU: Parametric Rectified Linear Unit).
The combination of BN and PReLU blocks allows a smoother representation in regression tasks than the combination with ReLU.
%
%
Finally, there is another convolutional layer with a ReLU, to reduce the number of channels to $1$ and obtain the enhanced log-spectrum for each signal.
Every WRB increases the number of channels used to get the outputs.
The widening operation is done in the first convolution of the first residual block of each WRB.
In order to compute the residual connection as a sum operation, the number of channels in the straight path and in the convolution path has to be the same.
Therefore, when the number of channels is increased, a Conv1D with $k=1$ is added.
This can be interpreted as a position wise fully connected layer to adjust the number of channels from the residual path to the number of channels in the convolutional path in order to add them.
In this work, we want to enhance the logarithmic spectrum of a noisy input signal $X_{Noise}$. 
For this objective we use the Mean Square Error (\emph{MSE}) in the training cost function to get an enhanced signal $X_{Enh}$ from $X_{Noise}$ as similar as possible to the clean reference $Y$.
From the experience in our previous work \cite{llombart2018wide}, Instead of using a frame by frame enhancement, we process the whole input signal as a sequence.
This means that instead of providing for each example the regression error of one frame, we propagate the accumulated error of the regression along the complete sentence. 
 This strategy considerably reduces the computation because instead of generating hundreds of training examples from one input signal, each training example is a complete input sequence.
Finally, the cost function is the mean of all input frames \emph{MSE} described in equation (\ref{eq:jm2m})
\begin{equation}
    J(Y, X_{Enh}) =\frac{1}{T}\sum^{T-1}_{t=0}\frac{1}{N}\sum^{N-1}_{n=0}MSE(y_{t,n}, x_{Enh,t,n})
 \label{eq:jm2m}
\end{equation}
where $T$ is the number of frames of the example, $N$ is the feature dimension, $y_{t,n}$ are $Y$ frames and $x_{Enh,t,n}$ are $X_{Enh}$ frames.

\section{Experimental setup}
\label{sec:exp}
%
The experimental framework developed in this work is inspired by the REVERB Challenge task\footnote{http://reverb2014.dereverberation.com}.
We evaluate the performance of speech enhancement methods through speech quality measures, aiming to find a trade-off between the dereverberation and the introduction of spectral distortion with the enhancement.
%

%
\subsection{Datasets}
%
Approaches were tested on the official Development and Evaluation sets of the REVERB Challenge \cite{Challenge2013}. 
The dataset has simulated speech from the convolution of WSJCAM0 Corpus \cite{WSJCAMO} with three measured Room Impulse Responses (RIR) ($RT_{60} = 0.25, 0.5, 0.7 s$) at two speaker-microphone distances: far (2 m) and near (0.5 m).
It was also added stationary noise recordings from the same rooms (SNR = 20 dB). Besides, it has real recordings, acquired in a reverberant meeting room ($RT_{60} = 0.7 s$) at two speaker-microphone distances: far (2.5 m) and near (1 m) from the MC-WSJ-AV corpus \cite{MC-WSJ-AV}.
We also used real speech samples from VoiceHome v0.2 \cite{VoiceHome1} and v1.0 \cite{VoiceHome2}.
VoiceHome was recorded in a real domestic environment, such that the background noise is that typically found in households e.g. vacuum cleaner, dishwashing or interviews on television.
For training the DNN we used 16 kHz sampled data from the following datasets: Timit \cite{garofolo1993darpa}, Librispeech \cite{panayotov2015librispeech}, and Tedlium \cite{rousseau2014enhancing}.
This data was augmented by adding artificially generated RIR ($RT_{60} = 0.05-0.8 s$), stationary and non-stationary noises from Musan dataset \cite{musan2015}, SNR = 5-25 dB, including music and speech, and scaling the time axis at the feature level.
%

%
\subsection{Methods for comparison}
We compare the performance of the proposed WRN speech enhancement method with the state-of-the-art dereverberation method called Weighted Prediction Error (WPE), which is known to effectively reduce reverberation in the framework of the REVERB dataset \cite{Kinoshita2017}. 
We used the more recent version of WPE\footnote{\url{https://github.com/fgnt/nara_wpe}} which is also based on DNN \cite{Kinoshita2017}.
However, WPE uses an architecture based on LSTM, which also provides us the possibility for comparing the speech enhancement solutions from the DNN architecture point of view. 
%

%
\subsection{Performance assessment}
%
The speech quality was measured in terms of the distortion introduced by the enhancement process by means of the Log-likelihood ratio\footnote{Originally known as Itakuta distance} (LLR) \cite{Loizou2011}.
This was computed in the active speech segments (determined by a Voice Activity Detection (VAD) algorithm \cite{VADLTSD}).
For this measure, the closer the target to the reference, the lower the spectral distortion, therefore smaller values indicate better speech quality.

On the other hand, we assess the reverberation level of the signal through the Speech-to-reverberation modulation energy ratio (SRMR) \cite{SRMRmeasure}. 
In this case, higher values indicate better speech quality.
Note that only SRMR can be used with real data because LLR is computed using the observed/enhanced signal and clean reference.
%

%
\subsection{Network configuration}
%
The front-end starts segmenting speech signals in 25, 50, and 75 ms Hamming-windowed frames, every 10 ms.
We provide this multiple representations of the input in order to maintain as much reverberant impulsive response inside the Hamming window as it is possible, without losing temporal resolution of the acoustic events.
For each frame segment, three types of acoustic feature vectors are computed and stacked, to create a single input feature vector for the network: 512-dimensional FFT, 32, 50, 100-dimensional Mel filterbank, and cepstral features (same dimension of the corresponding filterbank).
Finally, each feature vector is normalized by variance. 
Input features were generated and augmented on-the-fly, operating in contiguous vector blocks of 200 samples so that convolutions in the time axis can be performed.
The network uses four WRN blocks with a widen factor of 8.
AdamW algorithm was used to train the network and PReLUs \cite{He2015} as parametric non-linearity.
%

\section{Results and discussion}
\label{sec:res}
%
\begin{figure}[h]
\centerline{\includegraphics[width=1\linewidth]{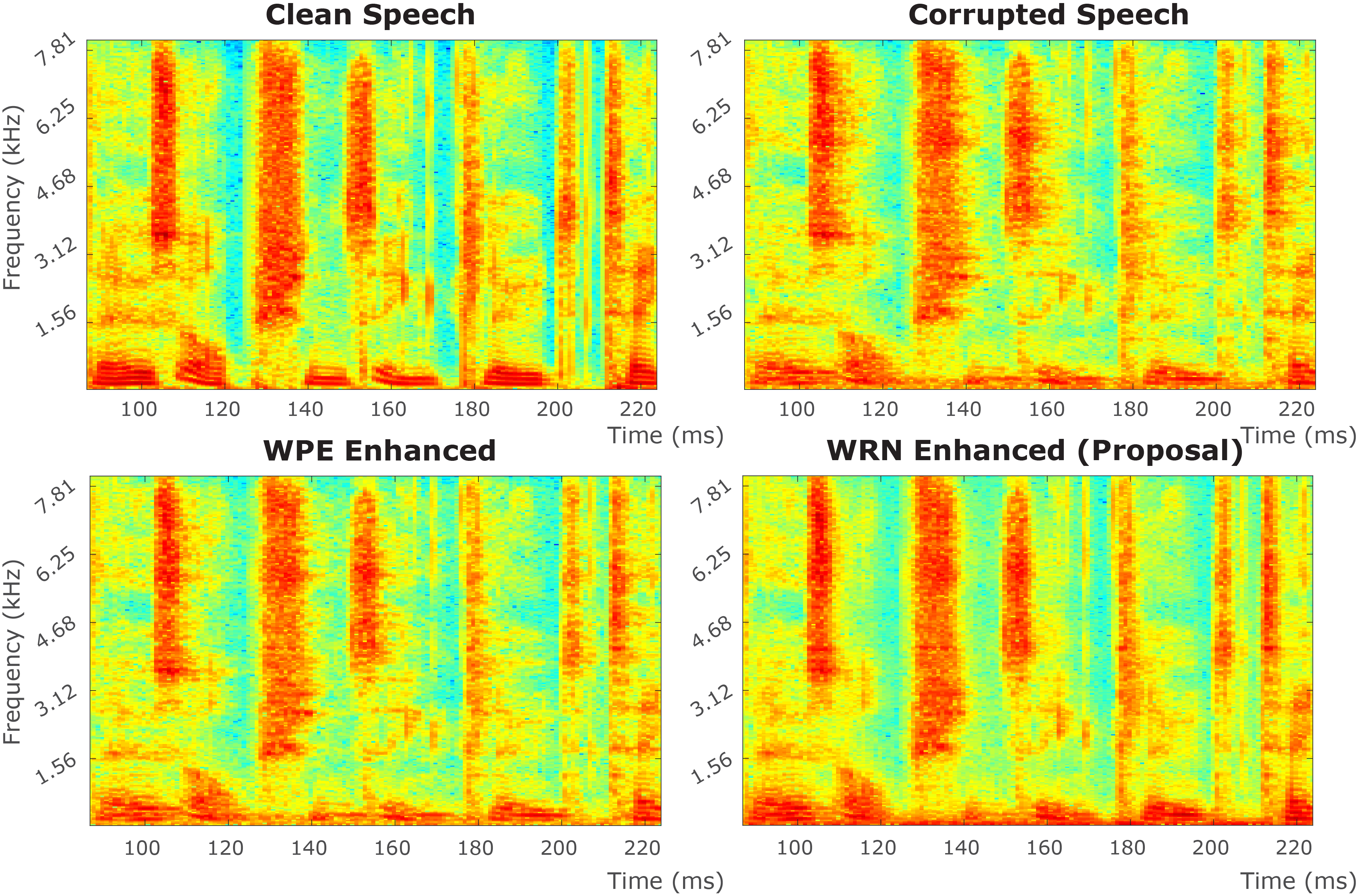}}
\caption{{\it Qualitative example of the enhancement performance in a single signal of REVERB Dev dataset.}}
\label{fig:goodresult}
\end{figure}

Figure \ref{fig:goodresult} shows a qualitative example of the enhancement performance in the signal c31c0204.wav of the REVERB Dev dataset ($RT_{60}$ = 0.25s, Distance speaker-mic = 200 cm). 
Observe the distortion due to reverberation in the corrupted speech spectrogram at the top-right side of the figure. Reverberation provokes a remarkable temporal spread of the power spectrum in active speech segments.  
Note that the enhanced speech through WPE removes some of this effect, but the WRN method is more accurate in this aim, achieving a better reconstruction of the signal. 
%

%
\subsection{Speech quality for processing tasks}

Table \ref{tabdistortion} presents speech quality results in terms of distortion with the LLR distance for simulated speech samples.
The first row corresponds to the reverberant unprocessed speech, which is compared to the quality achieved by the enhanced signals using WPE and the proposal WRN enhancement method.
Both DNN-based methods are able to enhance the corrupted speech data, but the proposal outperforms WPE in terms of spectral distortion. 
\begin{table} [h!]
\footnotesize
\caption{\label{tabdistortion} {\it LLR distance in simulated reverberated speech samples from REVERB Dev \& Eval datasets.}}
\centerline{
\begin{tabular}{c||c|c}
Methods & REV-Dev & REV-Eval \\  
\specialrule{.2em}{.1em}{.1em} 
Unprocessed & 0.63 & 0.64 \\ \hline       
WPE \cite{NaraWPE} & 0.60 & 0.60 \\ \hline 
WRN & \cellcolor{gray!35}0.50 & \cellcolor{gray!35}0.51 \\ 
\end{tabular}
}
\end{table}
%

\subsection{Speech quality for dereverberation: Simulated vs. Real}
%
Table \ref{resultsSRMR} shows the average of SRMR results over the evaluated conditions for simulated and real speech samples. 
The first column on the left corresponds to the unprocessed speech data.
Shadowed cells highlight the best results for each dataset.
\begin{table} [h!]
\caption{\label{resultsSRMR} {\it Speech quality through SRMR results for simulated and real reverberated speech samples.}}
\centerline{
\footnotesize
\begin{tabular}{c||c|c|c}
Datasets & Unprocessed & WPE \cite{NaraWPE} & WRN \\ \specialrule{.2em}{.1em}{.1em} 
\multicolumn{4}{c}{\textbf{Simulated}} \\ \hline 
REVERB Dev & 3.67 & 3.90 & \cellcolor{gray!35}4.75 \\ \hline 
REVERB Eval & 3.68 & 3.91 & \cellcolor{gray!35}4.63 \\ \hline 
\multicolumn{4}{c}{\textbf{Real}} \\ \hline 
REVERB Dev & 3.79 & 4.17 & \cellcolor{gray!35}4.79 \\ \hline 
REVERB Eval & 3.18 & 3.48 & \cellcolor{gray!35}4.20 \\ \hline 
VoiceHome v0.2 & 3.19 & 3.28 & \cellcolor{gray!35}5.03 \\ \hline 
VoiceHome v1.0 & 4.51 & 4.96 & \cellcolor{gray!35}5.92 \\ 
\end{tabular}
}
\end{table}

The proposal outperforms baselines for all datasets evaluated.
The consistency in performance through different datasets supports the robustness of the method. 
This indicates that its parameters are not adjusted to some specific set of speech signals, which is a desirable quality for an enhancement method.
These positive results beyond simulations, encourage the use of the method in realistic scenarios.
Furthermore, note the WRN model was trained with artificially synthesized reverberation, however, it showed to be effectively dealing with a reverberated speech from real-world scenarios. 
%

\subsubsection{Room sizes and reverberation level}
\begin{figure}[h!]
\centerline{\includegraphics[width=0.75\linewidth]{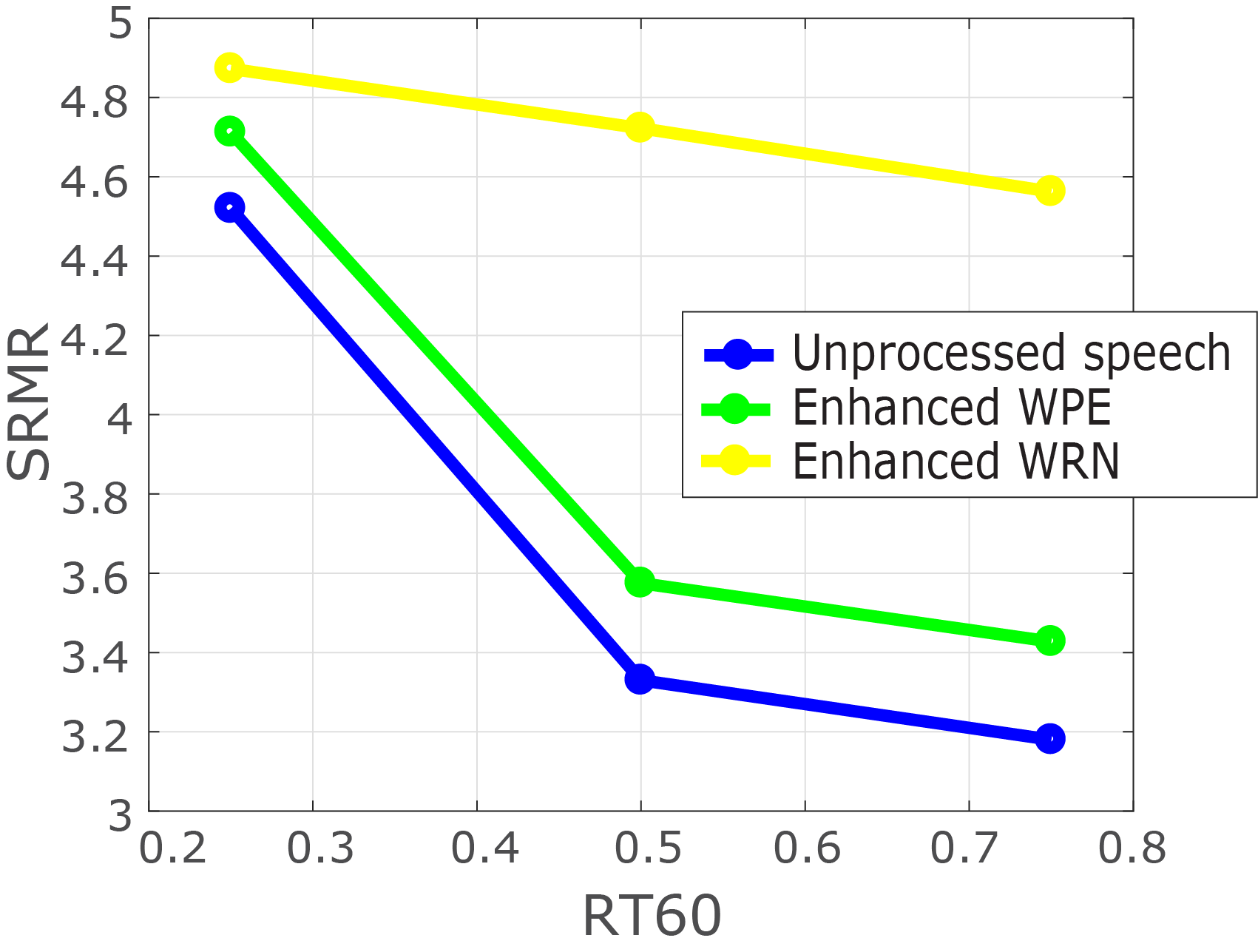}}
\caption{{\it Speech quality through SRMR measure for different reverberation levels in simulated reverberated speech samples from REVERB Dev \& Eval datasets.}}
\label{fig:RT60}
\end{figure}

%
Figure \ref{fig:RT60} shows the evolution of SRMR results with the increase of reverberation level for different room sizes: $Room1- RT_{60}=0.25s$, $Room2-RT_{60}=0.5s$, and $Room3-RT_{60}=0.75s$.
The proposed WRN method achieves higher speech quality than the reference for all conditions evaluated.
Furthermore, the results of the proposed method have less variability through the $RT_{60}$, indicating the robustness of the method in different scenarios.   
See that the speech quality improvement with respect to the reference methods increases with the $RT_{60}$. 
However, note that there is less space for improvement in the $Room1-RT_{60}=0.25s$ condition, so it is harder to enhance. 
%


\subsubsection{Near and Far field}
\begin{figure}[h!]
\centerline{\includegraphics[width=0.75\linewidth]{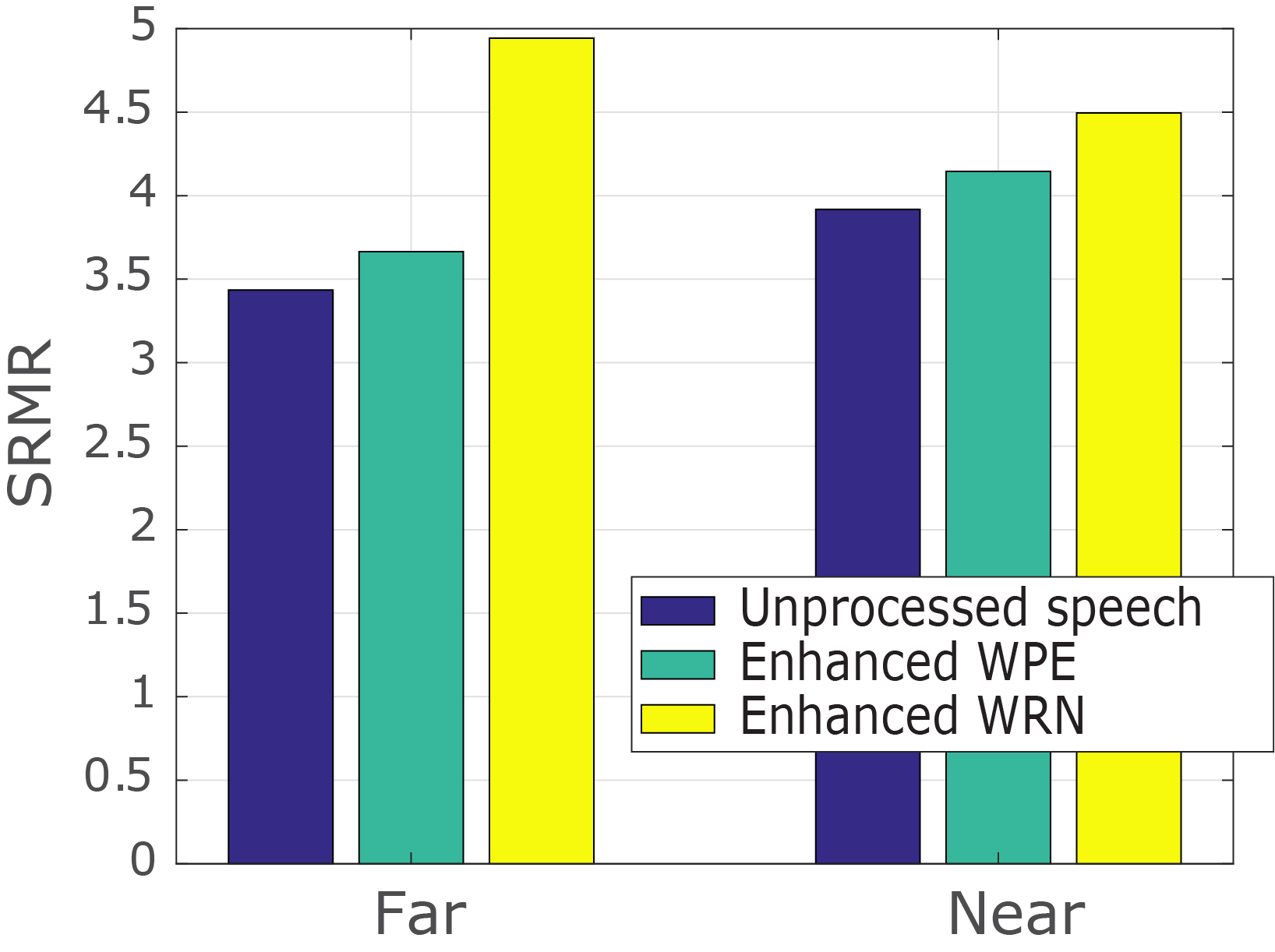}}
\caption{{\it SRMR results for near- and far-field simulated reverberated speech from REVERB Dev \& Eval datasets.}}
\label{fig:nearfar}
\end{figure}

%
Figure \ref{fig:nearfar} presents an average of SRMR results for $far$ (250 m) and $near$ (50 m) conditions in the simulated REVERB Dev and Eval datasets. 
WRN considerably outperformed the WPE baseline for $34,88 \%$ in far-field and $8,44 \%$ in near-field.
Note that the proposal is strongest in the far-field conditions, which is usually the most challenging scenario.    

\subsection{Effect of train-test mismatch}
%
As we saw before, the conditions for $Room1, RT_{60} = 0.25$ and $near$ speaker-microphone distance were harder to enhance. 
These cases correspond to low reverberation, where there is a small margin for improvement. 
Hence, to increase the focus of the data augmentation for the network training to these conditions could provide a boost of performance.   
Furthermore, note that due to the lack of exact room size values in the test dataset description, WRN data training included a reasonable estimation of small room size. 
However, this is probably not small enough for $Room1$. 
On the other hand, the distance configuration of the training data design considers the speaker/microphone can be randomly situated all-around the room, modeling it with uniform data distribution.
This left low probability for the specific test data distances of $near$ (50 cm) and $far$ (250 cm).
In order to improve the results in these scenarios, the training data in future approaches should include smaller room sizes, change the function which models the speaker-microphone distances, in order to increase the probability of certain distances.
However, caution should be taken with overfitting of the training data to some specific dataset.
A better compromise between the network generalization and the test data characteristics will be a more reasonable solution. 

\subsection{Comparison with previous work}
Experimental results evidenced that the proposed WRN architecture outperformed the reference WPE. 
Despite the potentiality of the RNN-LSTM architecture used in WPE, the combination of CNN with residual connections in the proposal was able to obtain more expressive representations of the reverberant speech.
This structure performs the enhancement taking into account the full utterance through convolutions in all temporal domain of the signal, which is higher while deeper is the structure. 
As WPE is based on RNN-LSTM it only takes into consideration previous context. However, for the enhancement purpose, a representation that considers a context including some future samples may contribute to increasing the overall performance. 
The proposed WRN architecture implements this idea through the convolutional layers. 

With our proposal, the reconstruction of the clean signal achieved improved speech quality more than the reference, with a proper trade-off between the level of dereverberation, and the amount of spectral distortion. 
These results were also validated through a test in real distorted speech, to show the generalization capability of the model.   

\section{Conclusions and future work}
\label{sec:conc}
This paper has introduced a novel speech enhancement method based on a WRN architecture that takes advantage of the powerful representations obtained from a wide topology of CNN with residual connections. Results showed the potentiality of the WRN providing an enhanced speech on top of the state-of-the-art RNN-LSTM-based method called WPE. Best results were obtained for far-field reverberated speech in three different room sizes. The residual mechanism was extremely useful in this case since the signal has always a linear shortcut and the non-linear path enhances it in certain steps by adding or subtracting corrections. In practical applications, this is a valuable property because realistic scenarios could challenge the system with many different conditions \cite{Ribas2016}. 

Despite results are encouraging the proposal can be further improved. Future work will focus on fine-tuning the data training configuration with a view to updating the compromise between generalization and accuracy. We also plan to expand the experimental setup to evaluate in speech recognition task with speech data in alternative scenarios from other datasets and comparative baselines. On the other side, the inclusion of perceptual features in the network cost function will be explored in order to improve the performance in the speech reconstruction process. 

\section{Acknowledgment}
Funded by the Government of Aragon (Reference Group T36\_17R) and co-financed with Feder 2014-2020 "Building Europe from Aragon".
This work has been supported by the Spanish Ministry of Economy and Competitiveness and the European Social Fund through the project TIN2017-85854-C4-1-R. We gratefully acknowledge the support of NVIDIA Corporation with the donation of the Titan Xp GPU used for this research. This material is based upon work supported by Google Cloud.

\bibliography{biblio.bib}
\bibliographystyle{IEEEtran}

\end{document}